\documentclass[12pt]{article}

\usepackage{amsmath, cite, graphicx}
\usepackage{multirow}
\usepackage{amsfonts}
\usepackage{amssymb}
\usepackage{amscd}
\usepackage{cite}
\usepackage{amsmath}
\usepackage{bbm}

\def\hybrid{\topmargin -20pt    \oddsidemargin 0pt
        \headheight 0pt \headsep 0pt
        \textwidth 6.25in       
        \textheight 9.5in       
        \marginparwidth .875in
        \parskip 5pt plus 1pt   \jot = 1.5ex}

\hybrid

\numberwithin{equation}{section}
\numberwithin{table}{section}

\newcommand\e{\mathrm{e}}
\newcommand\iu{\operatorname{i}}

\begin{document}

\begin{titlepage}

\begin{center}

\rightline{\small CERN-PH-TH/2015-212}

\vskip 2cm

{\Large \bf Supersymmetric branes on curved spaces and fluxes}

\vskip 1.7cm

{\bf Hagen Triendl} \\

\vskip 1.0cm

{\em Theory Division, Physics Department, CERN, CH-1211 Geneva 23, Switzerland}

\vskip 1.6cm

{\tt hagen.triendl@cern.ch} \\

\end{center}

\vskip 2cm

\begin{center} {\bf ABSTRACT } \end{center}

We discuss general supersymmetric brane configurations in flux backgrounds of string and M-theory and derive a necessary condition for the worldvolume theory to be supersymmetric on a given curved manifold.
This condition resembles very much the conditions found from coupling a supersymmetric field theory to off-shell supergravity but can be derived in any dimension and for up to sixteen supercharges. Apart from the topological twist, all couplings appearing in the supersymmetry condition are linked to fluxes in the bulk.
We explicitly derive the condition for D3-, M2- and M5-branes, in which case the results are also useful for constructing holographic duals to the corresponding field theories.
In $N=1$ setups we compare the supersymmetry conditions to those that arise by coupling the field theory to off-shell supergravity. We find that the couplings of both old and new minimal supergravity are simultaneously realized, indicating that off-shell supergravity should be coupled via the S-multiplet of 16/16 supergravity in order to describe all supersymmetric brane theories on curved spaces.

\vfill

\today

\end{titlepage}


\section{Introduction}

Brane physics are an important and powerful ingredient in many models of string and M-theory, as they introduce field theory sectors into a UV-complete theory. In many of these setups supersymmetry ensures control over the theory and enables to compute physical couplings. Naturally these branes live on curved space(-time)s and generically their worldvolume theories are coupled to the ambient geometry and fluxes. However, in the limit where the directions transverse to the brane become infinitely large one can decouple the worldvolume theory from the bulk dynamics, effectively sending the Planck scale on the worldvolume to infinity. In this limit the background fields are degraded to non-dynamical background fields that effectively induce certain couplings on the worldvolume of the brane. A natural question to ask is which supersymmetric field theories can be reproduced from a system of (intersecting) branes.

The most classic example of a coupling induced by the bulk theory is the curvature of the worldvolume that is induced by the bulk metric. It has for instance been shown a long time ago \cite{Bershadsky:1995qy} that D-branes wrapping a calibrated cycle within a Calabi-Yau manifold are described by a topologically twisted gauge theory \cite{Witten:1988ze,Yamron:1988qc,Vafa:1994tf}. This fits with the observation that in four dimensions an $N=1$ topologically twisted field theory can be constructed on any K\"ahler manifold \cite{Johansen:1994aw}, as calibrated four-cycles inside a Calabi-Yau threefold are indeed K\"ahler manifolds. $N=2$ topologically twisted theories in contrast can be constructed on any four-dimensional manifold \cite{Witten:1994ev}, which suggests that calibrated four-cycles inside a higher-dimensional supersymmetric gravitational string background might not have to be K\"ahler.

In the last years a lot of work has gone into constructing supersymmetric field theories on more general curved backgrounds \cite{Adams:2011vw}-\cite{Lawrence:2015qra}. This is usually done by coupling the field theory to an off-shell supergravity formulation and treating the supergravity fields as background fields that do not satisfy any equations of motion. In four dimensions usually there are two off-shell formulations used: the old minimal \cite{Stelle:1978ye,Ferrara:1978em} and the new minimal \cite{Sohnius:1981tp,Sohnius:1982fw} formulation. Both are special cases of the 16/16 supergravity \cite{Girardi:1984vq}-\cite{Siegel:1986sv}.
On a given curved manifold with a specific metric a field theory can be made supersymmetric by tuning the other, auxiliary fields in the supergravity multiplet. It has, in the Euclidean case for instance, been shown in \cite{Dumitrescu:2012ha} that by coupling the $N=1$ supersymmetric field theory to new minimal supergravity it can be made supersymmetric on any complex manifold. Coupling to old minimal supergravity also enables for instance the four-sphere, which does not even allow for an almost complex structure, to admit supersymmetric field theories \cite{Festuccia:2011ws}.

Off-shell supergravity has unfortunately various limitations. While there are multiple formalisms for off-shell supergravity with four supercharges, in the case with eight supercharges the coupling to arbitrary multiplets has not been worked out and in the case with sixteen supercharges the action is altogether unknown. This raises the question whether there is a another framework to discuss supersymmetric field theories on curved backgrounds. Also, so far it has not been generally discussed which of the supersymmetric field theories that arise in the constructions mentioned above are realized in terms of dynamical branes in string theory. One goal of this paper is to close this gap.

In the presence of fluxes, generically the bulk geometry is not Calabi-Yau any more (see \cite{Grana:2005jc} for a review) and in turn also brane calibration conditions and thereby the geometry of calibrated cycles are altered by the presence of fluxes \cite{Koerber:2005qi,Martucci:2005ht,Martucci:2011dn}. In this work we will link these observations to the above discussion of supersymmetric field theories on curved manifolds. For this, we will derive the differential conditions that govern the possibility of supersymmetry on the worldvolume of the brane. For a brane to be calibrated, the bulk background must firstly be supersymmetric. This is determined by the supersymmetry variations of the fermions in the ten- or eleven-dimensional supergravity. The brane calibration condition can then conveniently be expressed in terms of kappa symmetry on the brane world volume \cite{Cederwall:1996ri,Bergshoeff:1996tu}. Combining these two conditions gives a number of conditions on the embedding of the brane and a differential equation for spinors on the worldvolume. While the former is model-dependent and governs the embedding of the brane into the ambient geometry, the latter gives a supersymmetry condition for the curved space itself that resembles the condition found in off-shell supergravity. We will show how to derive this supersymmetry condition for various brane configurations.
Our most prominent example will be a stack of D3-branes. Its supersymmetry condition is derived in \eqref{eq:susy4dN=4}. Similarly, the supersymmetry condition for a stack of M2- and for a stack of M5-branes is derived in \eqref{eq:susyM2} and \eqref{eq:susyM5}, respectively.

The supersymmetry condition only depends on the geometry close to and the form field strengths of the backgrounds at the location of the brane. The geometry close to the brane determines the worldvolume geometry and the R-symmetry twist. The form field strengths give rise to additional terms in the supersymmetry condition and resemble the auxiliary fields of off-shell supergravity. The identification of the various couplings can be found for instance in \eqref{eq:susy4dN=4def} for D3-branes. In all cases, the worldvolume couplings for the field theory are completely determined by the fluxes in the ambient background.

The condition for a the field theory on a stack of branes to be supersymmetric naturally appears in an $N=4$ version, i.e.\ the spinor appearing in this equation has sixteen components. The topological twist is embedded into the respective R-symmetry group and also the other couplings appear in representations of that group. By considering the stack of branes to lie inside or intersect with other brane stacks additional kappa-symmetry projections are applied in order to reduce the number of spinor components and the R-symmetry of the theory. In this way it is also possible to make contact with the supersymmetry conditions coming from the various off-shell supergravity formulations. We consider for instance a stack of D3-branes lying inside the intersection of D7-branes and derive the condition for the theory to be supersymmetric in \eqref{eq:susy4dN=1}. In this case, four supercharges appear in the supersymmetry condition, and therefore it can be compared to the condition appearing in \cite{Festuccia:2011ws,Dumitrescu:2012ha}. Indeed we will show that the possible couplings on the brane induced by fluxes \eqref{eq:susy4dN=1def} encompass the couplings from both old and new minimal supergravity. In particular, old minimal supergravity is the special case of \eqref{eq:susy4dN=1} for $A=0$, while new minimal supergravity is represented by the case $M=\bar M=0$. The generic case is described by coupling the field theory to off-shell 16/16 supergravity, via the S-multiplet of \cite{Komargodski:2010rb}, which is the most general way of doing so \cite{Dumitrescu:2011iu}.\footnote{We would like to thank C.\ Closset very much for pointing out this connection.}
Unfortunately an classification of supersymmetric field theories on curved backgrounds using this formalism has not been attempted yet.

In this work we will mostly focus on D3-, M2- or M5-branes in curved backgrounds, as they play a prominent role in the construction of holographic AdS/CFT pairs. We hope that the presented results will also be useful to understand the holographic dual gravitational background. We suspect that in the t'Hooft limit any system of supersymmetric D3-, M2- or M5-branes give rises to a supersymmetric field theory with a holographic dual. Translating the bulk fields into worldvolume couplings therefore automatically leads to a dictionary between the couplings of field theory and the corresponding field configurations in ten- or eleven-dimensional supergravity and thus in AdS spacetime, enabling a more systematic study of these field theories holographically.

Also note that under dimensional reduction from M-theory to type IIA, the supersymmetry conditions do not change. Therefore our results automatically carry over to the case of field theories on D2- and D4-branes (as well as NS5-branes and F-strings of type IIA).

The paper is structured as follows. In Section \ref{sec:general} we discuss how the supersymmetry condition in the bulk and kappa symmetry for the brane together give a differential condition on the brane worldvolume. In Section \ref{sec:D3branes} we then study field theory on D3-branes in more detail and make contact with the formulations of old and new minimal supergravity. In Section \ref{sec:membranes} we discuss supersymmetric theories on M2- and M5-branes, and in Section \ref{sec:examples} we present some simple examples. A discussion of our findings can be found in Section \ref{sec:conclusion}.

\section{Supersymmetric string backgrounds and branes}
\label{sec:general}

\subsection{Supersymmetry and calibration conditions}

We will assume a general supersymmetric background of type II or M-theory. In type II the supersymmetry transformation of the spinor doublet $\Psi_M$ in the democratic formulation reads
\begin{equation}\label{eq:gravitinovar}
\delta_\varepsilon \Psi_M =  \nabla_M \varepsilon + \tfrac18 H_{MNP} \Gamma^{NP} {\cal P} \varepsilon + \tfrac{1}{16} \e^{\phi} \sum_n \tfrac{1}{(2n)!} F_{M_1 \dots M_{2n}} \Gamma^{M_1 \dots M_{2n}} \Gamma_M {\cal P}_{2n} \varepsilon \ ,
\end{equation}
and the dilatino variation is
\begin{equation}
 \delta_\varepsilon \lambda = ( (\partial_M \phi)\Gamma^M +\tfrac16 H_{MNP} \Gamma^{MNP}{\cal P}) \varepsilon + \tfrac18 \e^{\phi} \sum_n (-1)^{2n} \tfrac{5-2n}{(2n)!}  F_{M_1 \dots M_{2n}} \Gamma^{M_1 \dots M_{2n}} {\cal P}_{2n} \varepsilon \ ,
\end{equation}
where
\begin{equation}
{\cal P} = \left(\begin{aligned} 1 && 0 \\ 0 && -1 \end{aligned}\right) \ , \qquad {\cal P}_n = \left(\begin{aligned} 0 && 1 \\  (-1)^{[(n+1)/2]} && 0 \end{aligned}\right) \ ,
\end{equation}
the number $n$ runs over integers (integers plus $1/2$) in type IIA (IIB) and $[\cdot ]$ denotes the floor function, i.e.\ $[(n+1)/2]$ is $n/2$ in type IIA and $(n+1)/2$ in type IIB. Note that $\cal P$, ${\cal P}_n$ and ${\cal P}_{n+2}$ anticommute with each other and obey the commutation relations
\begin{equation} \label{eq:Pn_comm}
 [{\cal P}_n, {\cal P}] = 2 {\cal P}_{n+1} \ , \qquad [{\cal P}_n, {\cal P}_m] = ((-1)^{[n+1/2]} - (-1)^{[m+1/2]} ) {\cal P} \ .
\end{equation}
Furthermore, in type IIB the two spinors in the doublet $\varepsilon= (\varepsilon^1, \varepsilon^2)$ are of the same chirality, which means that the ${\cal P}_n$ and ${\cal P}$ commute with the spacetime gamma matrices and generate ${\rm Sl}(2, \mathbb{R})$ of type IIB. In type IIA the two spinors in the doublet $\varepsilon= (\varepsilon^1, \varepsilon^2)$ have opposite chirality, indicating that ${\cal P}_n$ anti-commutes with the chirality operator $\Gamma_{(10)} = \tfrac{1}{10!} \epsilon^{M_0\dots M_{9}} \Gamma_{M_0\dots M_{9}}$. Indeed $\Gamma_{(10)}{\cal P}$ is the identity, and the matrices $\tilde \Gamma_{\tilde M}$, $\tilde M=0,\dots,10$, defined by
\begin{equation} \label{eq:duality_MIIA}
\tilde \Gamma_{M}=\Gamma_{M} {\cal P}_0 \ , \qquad \tilde \Gamma_{10} = {\cal P} \ ,
\end{equation}
form the eleven-dimensional Clifford algebra.\footnote{Note that in the entire paper we will use flat indices to keep the formulas as simple as possible.} This is of course natural as type IIA can be lifted to M-theory.

In the following we will discuss branes in supersymmetric backgrounds. These backgrounds are distinguished by the fact that all fermion supersymmetry variations vanish, which means that
\begin{equation}\label{eq:gravvarzero0}
\delta_\varepsilon \Psi_M = 0 \ ,
\end{equation}
and
\begin{equation}\label{eq:dilvarzero}
\delta_\varepsilon \lambda = 0 \ .
\end{equation}
Note that while \eqref{eq:dilvarzero} determines the dilaton profile, \eqref{eq:gravvarzero} is a condition on the ten-dimensional background.
In the following we will write \eqref{eq:gravvarzero0} as
\begin{equation}\label{eq:gravvarzero}
D_M \varepsilon = 0 \ ,
\end{equation}
where we defined the differential operator $D_M$ acting on the spinor bundle as
\begin{equation}\label{eq:defD}
D_M  = \nabla_M  + \tfrac18 H_{MNP} \Gamma^{NP} {\cal P} + \tfrac{1}{16} \e^{\phi} \sum_n \tfrac{1}{(2n)!} F_{M_1 \dots M_{2n}} \Gamma^{M_1 \dots M_{2n}} \Gamma_M {\cal P}_{2n} \ .
\end{equation}

A supersymmetric brane on a $(p+1)$-dimensional cycle ${\cal S}_{p+1}$ inside the string background can be described by the kappa symmetry condition
\begin{equation} \label{eq:kappa_symm}
 \hat \Gamma \hat \varepsilon  = \hat \varepsilon \ .
\end{equation}
where we defined $\hat \varepsilon$ as the restriction of $\varepsilon$ to ${\cal S}$, i.e.\
\begin{equation}
\hat \varepsilon = \varepsilon \Big|_{\cal S} \ .
\end{equation}
For a D$p$-brane in type II string theory $\hat \Gamma$ is given by
\begin{equation} \label{eq:kappa_projector}
\hat \Gamma_{{\rm D}p} = (-{\rm det}(g+ {\cal F}))^{-1/2}\sum_{2l+s = p+1} \tfrac{1}{l! s! 2^l} \epsilon^{n_1\dots n_{2l} m_{1} \dots m_s } {\cal F}_{n_1 n_2} \dots {\cal F}_{n_{2l-1} n_{2l}} \gamma_{m_1 \dots m_s} {\cal P}_{s+1} \ ,
\end{equation}
where $\gamma$ are the anti-symmetric products of gamma matrices on the worldvolume, which are related to the anti-symmetric product of $\Gamma$ matrices by the push-forward $i_*$ of the embedding function $i$ of ${\cal S}$ into $M$.
Here we assumed that the D$p$-brane fills out the time direction, which means that $\gamma_0$ is one of the gamma matrices in the anti-symmetric product. If the D$p$-brane does not fill out time, $\hat \Gamma_{{\rm D}p}$ has an additional factor of $\iu$, and there is no minus sign in front of the determinant.
In the following we will use \eqref{eq:kappa_projector} in flat coordinates so that $g$ is just the flat metric.

We can perform a similar analysis in M-theory. In M-theory there is only one eleven-dimensional gravitino $\Psi_M$ and one eleven-dimensional supersymmetry parameter $\varepsilon$. The supersymmetry variation of the gravitino here reads
\begin{equation}
\delta_\varepsilon \Psi_M = \nabla_M \varepsilon - \tfrac{1}{12}( \tfrac{1}{4!} G_{NPQR} \Gamma^{NPQR}{}_M - \tfrac{2}{3!} G_{MNPQ} \Gamma^{NPQ}) \varepsilon \ .
\end{equation}
In terms of \eqref{eq:gravvarzero} we can define the differential operator
\begin{equation}\label{eq:defDM}
D_M = \nabla_M - \tfrac{1}{12}( \tfrac{1}{4!} G_{NPQR} \Gamma^{NPQR}{}_M - \tfrac{2}{3!} G_{MNPQ} \Gamma^{NPQ}) \ ,
\end{equation}
analogous to \eqref{eq:defD}.
The kappa symmetry projector for an M5-brane with worldvolume flux ${\cal H}$ reads in flat coordinates
\begin{equation}\label{eq:kappaM5}
\hat \Gamma_{{\rm M}5} = c({\cal H})^{-1/2} \epsilon^{m_{1} \dots m_6 } ( \tfrac{1}{6!} \gamma_{m_1 \dots m_6} + \tfrac{1}{36} {\cal H}_{m_1 m_2 m_3} \gamma_{m_4 m_5 m_6} + \tfrac{1}{8} {\cal H}_{m_1 m_2}{}^m {\cal H}_{m_3 m_4 m} \gamma_{m_5 m_6} ) \ ,
\end{equation}
where
\begin{equation}
c({\cal H}) = 1 - \tfrac{1}{6} {\cal H}_{mnp} {\cal H}^{mnp} - \tfrac{1}{4!}{\cal H}_{[mn}{}^r {\cal H}_{p q] r} {\cal H}^{mns} {\cal H}^{pq}{}_s \ .
\end{equation}
Similarly, we have for an M2-brane
\begin{equation}\label{eq:kappaM2}
\hat \Gamma_{{\rm M}2} = \gamma_{(3)}=\tfrac{1}{3!} \epsilon^{m_{1} \dots m_3 } \gamma_{m_1 \dots m_3}  \ .
\end{equation}
There are also less-discussed 9- and 6-branes in M-theory \cite{Hull:1997kt}. The former has been dubbed M9-brane in \cite{Bergshoeff:1998bs}. We will later use them to reduce supersymmetry for the theory on the worldvolume of M5- and M2-branes. For this we use the projectors
\begin{equation} \label{eq:Mbrane_exotic}
\hat \Gamma_{{\rm M}9} = \tfrac{1}{10!}  \epsilon^{m_{1} \dots m_{10} }\gamma_{m_1 \dots m_{10}} \ , \qquad
\hat \Gamma_{{\rm M}6} = \tfrac{1}{7!} \epsilon^{m_{1} \dots m_{7} }  \gamma_{m_1 \dots m_{7}} \ .
\end{equation}
Also the M-theory kappa symmetry projectors have an additional factor of $\pm \iu$ in case they do not fill out the time direction.
By dimensional reduction the M-theory analysis gives also the correct type IIA result, as the identification \eqref{eq:duality_MIIA} suggests.

\subsection{Supersymmetry on the curved worldvolume}
For a supersymmetric brane in a supersymmetric background we have two supersymmetry conditions on the brane worldvolume $\cal S$: the kappa symmetry condition \eqref{eq:kappa_symm} and the restriction of \eqref{eq:gravvarzero} to $\cal S$. The latter splits on $\cal S$ into two components, one along the brane worldvolume and one normal to it. The component normal to the brane describes how $\hat \varepsilon$ is continued into the bulk using parallel transport and only gives a condition on the embedding of the brane into the supersymmetric background. The component tangential to the worldvolume gives a differential equation for $\hat \varepsilon$ itself, given by\footnote{We will use in the following flat coordinates, with the index $m$ ($a$) describing directions along (perpendicular to) the brane.}
\begin{equation} \label{eq:Depsilon0}
\hat D_m \hat \varepsilon = 0 \ ,
\end{equation}
where the operator $\hat D_m$ is given by $\hat D_m = D_m \Big|_{\cal S}$.
This means that both \eqref{eq:kappa_symm} and \eqref{eq:Depsilon0} are conditions on $\hat \varepsilon$ itself.
Under assuming the kappa symmetry projection \eqref{eq:kappa_symm}, we will study in the following the equation \eqref{eq:Depsilon0} in the equivalent form
\begin{equation} \label{eq:commkappa}
[\hat D_m, \hat \Gamma] \hat \varepsilon = 0 \ ,
\end{equation}
and
\begin{equation} \label{eq:anticommkappa}
 \{ \hat D_m, \hat \Gamma \} \hat \varepsilon = 0 \ .
\end{equation}

Since the ten-dimensional tangent space splits on the worldvolume according to
\begin{equation} \label{eq:tangentsplit}
TM \big|_{\cal S} = T {\cal S} \oplus N {\cal S} \ .
\end{equation}
the Lorentz group also breaks into two subgroups, acting on the tangent and normal bundle of $\cal S$, respectively. The Lorentz group is broken to $O(1,p) \times G_{\rm R}$, where $G_{\rm R}$ is $O(9-p)$ in type II and $O(10-p)$ in M-theory.\footnote{For an Euclidean brane the Lorentz group is broken to $O(p+1) \times G_{\rm R}$ instead and $G_{\rm R}$ is $O(1,8-p)$ in type II and $O(1,9-p)$ in M-theory.}
While the former is the Lorentz group on $\cal S$, the latter also transforms spinors on the world-volume and can therefore be understood as the R-symmetry on $\cal S$. In this language $\hat \varepsilon$ is a spinor on $\cal S$ valued in a non-trivial representation of the R-symmetry group ${\rm Spin}(9-p)$ or ${\rm Spin}(10-p)$ , and both \eqref{eq:commkappa} and \eqref{eq:anticommkappa} give conditions on the supersymmetry generators on the worldvolume and therefore for $\cal S$ to accommodate for a supersymmetric field theory.

Equation \eqref{eq:commkappa} is an algebraic equation, since $\hat \Gamma$ commutes with the connection $\tilde \nabla$ on $\cal S$. Another algebraic equation comes from considering
\begin{equation}\label{eq:commkappaA}
[\hat D_a, \hat \Gamma]\hat \varepsilon = 0 \ ,
\end{equation}
where $\hat D_a = D_a \Big|_{\cal S}$.
The simplest way to solve this equation is by setting the matrices $[\hat D_m, \hat \Gamma] $ and $[\hat D_a, \hat \Gamma] $ to zero, which constraints the bulk fields at the position of the brane. Otherwise these matrices work as projectors on the possible supersymmetry generators $\hat \varepsilon$. We are in this work mainly interested in what are the constrains for a manifold $\cal S$ to admit supersymmetric theories, and thus we will ignore \eqref{eq:commkappa} and \eqref{eq:commkappaA} and just study \eqref{eq:anticommkappa} to understand the restrictions on $\cal S$.

In contrast \eqref{eq:anticommkappa} is a differential equation for $\hat \varepsilon$ on the worldvolume. We can decompose the operator $\{ \hat \nabla_m, \hat \Gamma\}$ where $\hat \nabla = \nabla \Big|_{\cal S}$ into the induced connection $\tilde \nabla = i^*(\nabla)$ on the worldvolume plus the spin connection part $(\Omega_m)_{ab}$ acting on the normal bundle, as the off-diagonal components of the connection do not appear in the anti-commutator. Thus for each $m$ the matrix $(\Omega_m)_{ab}$ generates rotations on the normal bundle, i.e.\ a subgroup $H_{\rm R}$ inside $G_{\rm R}$.
Moreover, $\hat \varepsilon$ will be in a certain representations $\chi^{i}$ of $H_{\rm R}$, and we can decompose $\hat \varepsilon $ as
\begin{equation} \label{eq:spinor_decomposition}
\hat \varepsilon = \sum_{i,\alpha} \chi^{i}_\alpha \otimes \eta^{i}_\alpha  \ ,
\end{equation}
where $\alpha$ runs over certain representations of $H_{\rm R} $, and we can define a connection $A$ so that
\begin{equation}\label{eq:toptwist}
 (A_m)_\beta{}^\alpha \chi_\beta = \tfrac12 (\Omega_m)_{ab} \hat \gamma^{ab}  \chi_\alpha \ ,
\end{equation}
where we denote by $\hat \gamma^a$ the gamma matrices of the Clifford algebra on $N {\cal S}$.
Thus $A$ is the connection for a gauged subgroup $H_{\rm R}$ of the R-symmetry group $G_{\rm R}$.
In total we find
\begin{equation} \label{eq:connWV}
\{ \hat \nabla_m, \hat \Gamma\} \hat \varepsilon  = \sum_\alpha \chi^i_\alpha \otimes (\tilde \nabla_m + A_m) \eta^i_\alpha  \ .
\end{equation}
The gauging of the group $H_{\rm R}$ inside the R-symmetry group $G_{\rm R}$ is called a topological twist \cite{Johansen:1994aw,Witten:1994ev}.
Note that depending on the dimension of $\cal S$ the Majorana (and for type II the Weyl) condition for $\varepsilon$ and kappa symmetry \eqref{eq:kappa_symm} both can imply non-trivial relations between the different components $\chi^i_\alpha \otimes \eta^i_\alpha$.

For branes in purely gravitational backgrounds, i.e.\ where the form field strengths are set to zero, the vanishing of the gravitino variation \eqref{eq:gravitinovar} simplifies to
\begin{equation} \label{eq:covconst}
\nabla_M \varepsilon = 0 \ ,
\end{equation}
and the ten-dimensional background admits at least one covariantly constant spinor.
We can use \eqref{eq:connWV} to rewrite \eqref{eq:covconst} on the worldvolume as
\begin{equation}
\tilde \nabla \eta + A \cdot \eta = 0 \ .
\end{equation}
Hence the field theory on the brane world volume is a supersymmetric theory with a topological twist given by $H_{\rm R}$. This reproduces the result of \cite{Bershadsky:1995qy}.

Note that the dilatino variation \eqref{eq:dilvarzero} leads to a differential equation for the dilaton in the background. If we combine \eqref{eq:dilvarzero} with \eqref{eq:kappa_symm}, we get from the (anti-)commutator of the operator acting on $\varepsilon$ a differential equation for the dilaton profile along (perpendicular to) the brane. The profile of the dilaton perpendicular to the brane only determines the embedding of the brane, and so only the commutator equation is relevant for the worldvolume theory. We will not further discuss it since it only determines the dilaton profile on $\cal S$. Note however that allowing a solution for the dilaton might restrict the geometry of $\cal S$.

The equations governing the embedding of the brane into the ambient geometry are the before-mentioned equation for the dilaton profile transverse to the brane and
\begin{equation} \label{eq:embedding}
 \{ D_a, \hat \Gamma \} \varepsilon = 0 \ ,
\end{equation}
which should be understood as an equation valid in the proximity of the brane.
This equation gives a differential equation to continue $\hat \varepsilon $ into the bulk. It can be used to embed a supersymmetric field theory on a brane into a full string background, for instance by using a $1/R$ expansion. We will not study this equation further but notice that \eqref{eq:embedding} can be used to construct a (holographic) string background for a given worldvolume theory.

\section{Curved D3-branes in flux backgrounds}
\label{sec:D3branes}
Let us now be more specific and discuss four-dimensional field theories on D3-branes. We will in the following ignore worldvolume fluxes on the brane and set them therefore to zero. It would be very interesting to generalize the following discussion to the case of non-zero worldvolume flux.

\subsection{A stack of D3-branes}
The preserved supersymmetry of a D3-brane without worlvolume fluxes is described by the kappa symmetry
\begin{equation}\label{eq:projD3}
\gamma_{(4)} {\cal P}_1 \hat \varepsilon = \hat \gamma_{(6)} {\cal P}_1 \hat \varepsilon = \hat \varepsilon\ ,
\end{equation}
where we used that $\hat \varepsilon$ is chiral in ten dimensions. The chirality operators in four and six dimensions are defined by $ \gamma_{(4)} = \tfrac{1}{4!} \epsilon^{mnpq} \gamma_{mnpq}$ and $\hat \gamma_{(6)} = \tfrac{1}{6!} \epsilon^{abcdef} \hat \gamma_{abcdef}$.
The supersymmetry condition \eqref{eq:anticommkappa} then can be computed to be
\begin{equation}\label{eq:susy4d} \begin{aligned}
(\tilde \nabla_m + \tfrac12 (\Omega_m)_{ab} \hat \gamma^{ab}  - \tfrac{1}{8} \e^{\phi} (F_n \gamma^n \gamma_{(4)}   + \tfrac{1}{48} \epsilon^{abcdef} F_{cdefn} \gamma^{n}   \hat \gamma_{ab} )\gamma_m  )\hat \varepsilon   & \\
+ (\tfrac14 (H_{mna} - \tfrac12 \e^{\phi} F_{mna} \gamma_{(4)} + \tfrac14\e^{\phi} \epsilon_{mn}{}^{pq} F_{pqa}) \gamma^{n} \hat \gamma^a
+ \tfrac{1}{48} \e^{\phi} F_{abc}  \hat \gamma^{abc}  \gamma_{(4)} \gamma_m  ) {\cal P} \hat \varepsilon & = 0 \ ,
\end{aligned} \end{equation}
where we used \eqref{eq:Pn_comm}, \eqref{eq:projD3} and the self-duality of $F_5$ in ten dimensions.

Let us now decompose the spinor $\hat \varepsilon $ into an $SO(1,3)$ spinor on $T \cal S$ and an $SO(6)$ spinor on the $N {\cal S}$. From the Majorana-Weyl condition and \eqref{eq:kappa_symm} we find
\begin{equation}
\hat \varepsilon  = \sum_\alpha \left( \begin{aligned}
\chi_\alpha^{+} \otimes \eta_\alpha^+ + \chi_\alpha^{-} \otimes \eta_\alpha^- \\
-\iu \chi_\alpha^{+} \otimes \eta_\alpha^+ + \iu \chi_\alpha^{-} \otimes \eta_\alpha^-
\end{aligned} \right) \ ,
\end{equation}
where the upper index indicates the chirality of the spinors, defined by
\begin{equation}
 \gamma_{(6)} \chi^{\pm} = \mp \iu \chi^{\pm} \ , \qquad \hat \gamma_{(4)} \eta^{\pm} = \pm \iu \eta^{\pm} \ .
\end{equation}
Note that in four-dimensional spacetime and in six-dimensional Euclidean space the spinors of opposite chirality are related by charge conjugation, in other words there are charge conjugation operators $C_{(4)}$ and $C_{(6)}$ such that
\begin{equation}
\chi^{-} = C_{(6)} \chi^+ \ , \qquad  \eta^{-} = C_{(4)} \eta^{+} \ .
\end{equation}
Now we can rewrite \eqref{eq:susy4d} in a four-dimensional form
\begin{equation}\label{eq:susy4dN=4} \begin{aligned}
(\tilde \nabla_m + A_m  + \iu s_n \gamma^n \gamma_m   + U_n \gamma^{n} \gamma_m  )\eta^+ + (V_{mn} \gamma^n + \iu W \gamma_m + \iu t \gamma_m)\eta^-  & = 0 \ , \\
(\tilde \nabla_m + A^c_m  - s_n \gamma^n \gamma_m    + U^c_n \gamma^{n} \gamma_m  )\eta^- + (V^c_{mn} \gamma^n - \iu  W^c \gamma_m + \iu t \gamma_m)\eta^+  & = 0 \ ,
\end{aligned} \end{equation}
where the connection $A$ of the topological $SU(4)$ twist is defined by \eqref{eq:toptwist} and the other couplings are the ${\rm su}(4)$ matrices $U_m,V^\pm_{mn}$ and $W$ as well as the ${\rm SU}(4)$ singlets $s_m$ and $t$. They are defined by
\begin{equation} \label{eq:susy4dN=4def} \begin{aligned}
(U_m)_\beta{}^\alpha \chi^+_\beta = &  - \tfrac{1}{384} \e^{\phi} \epsilon^{abcdef} F_{cdefm} \hat \gamma_{ab} \chi^+_\alpha \ , \\
(V_{mn})_\beta{}^\alpha \chi^+_\beta = & \tfrac14 (H_{mna} + \tfrac12 \iu \e^{\phi} F_{mna}  + \tfrac14\e^{\phi} \epsilon_{mn}{}^{pq} F_{pqa})\hat \gamma^{a} \chi^-_\alpha \ , \\
W_\beta{}^\alpha \chi^+_\beta+  t \chi^+_\alpha= & - \tfrac{1}{48} \e^{\phi} F_{abc}  \hat \gamma^{abc} \chi^-_\alpha \ , \\
s_m = & - \tfrac{1}{8} \e^{\phi} F_m  \ .
\end{aligned} \end{equation}

\subsection{$N=2$ and $N=1$ cases}
So far we discussed the supersymmetry on the worldvolume in a $N=4$ way. In particular, there is no matter on a single smooth stack of D3-branes. Matter is generated if we have for instance flavor D7-branes so that the D3-branes lie inside the D7-branes. Let us first consider just one stack of D7-branes of that type. From the D3-brane worldvolume point of view, this corresponds to an additional projection for $\hat \varepsilon$ imposed by kappa symmetry of the D7-branes
\begin{equation}
\hat \Gamma_{{\rm D}7}\hat \varepsilon = \hat \varepsilon \ ,
\end{equation}
where
\begin{equation}
\hat \Gamma_{{\rm D}7} = \gamma_{(4)} \hat \gamma_{4567} {\cal P}_1 \ .
\end{equation}
In practical terms, this is the projection
\begin{equation}\label{eq:N=2projection}
\hat \gamma_{4567} \chi_\alpha^\pm = \chi_\alpha^\pm \ .
\end{equation}
This breaks the symmetries of the normal bundle to ${\rm SO}(2)\times {\rm SO}(4)$. Moreover, it restricts the ${\rm SO}(4)$ representation of the $\chi_\alpha^\pm$ to the chiral representation, which only transforms under ${\rm SU}(2)\subset {\rm SO}(4)$. Thus, the R-symmetry is ${\rm U}(1)\times {\rm SU}(2)$ and the connection $(\Omega_m)_{ab} $ can only gauge a subgroup of it.

Following up on this discussion we can define the ${\rm U}(1)\times {\rm SU}(2)$ action via the generators $\sigma_{\hat \imath} = (1, \sigma_i)$ such that
\begin{equation}  \begin{aligned}
\hat \gamma_{89} \eta^+ = & -\iu \eta^+ \ ,\\
\tfrac{1}{2} (\hat \gamma_{47} + \hat \gamma_{56}) \chi^+ = & \iu \sigma_1 \chi^+ \ , \\
\tfrac{1}{2} (\hat \gamma_{46} - \hat \gamma_{57}) \chi^+ = & \iu \sigma_2 \chi^+ \ , \\
\tfrac{1}{2} (\hat \gamma_{45} + \hat \gamma_{67}) \chi^+ = & \iu \sigma_3 \chi^+ \ .
\end{aligned} \end{equation}
Here the $\sigma_i$ are the Pauli matrices. For convenience we will split in the following the index $a$ into $a = \hat a, 8,9$.
By considering the anti-commutator of \eqref{eq:susy4d} with the projection \eqref{eq:N=2projection} we can simplify \eqref{eq:susy4d} to
\begin{equation} \label{eq:susy4dN=2} \begin{aligned}
(\tilde \nabla_m + \iu A_m^{\hat \imath} \sigma_{\hat \imath}  + \tfrac{1}{8} \e^{\phi} ( F^n  + \iu F^{n\hat \imath}   \sigma_{\hat \imath}  )\gamma_{n} \gamma_m )\eta^{+}  & \\
+ (G_{mn} \gamma^{n} +  F^{i} \sigma_i \gamma_{m} + \tfrac12 \tilde G_{np}\gamma^{np}{}_{m}  ) C \cdot \eta^{-}  &  = 0 \ ,
\end{aligned} \end{equation}
where we defined
\begin{equation} \label{eq:susyparN=2} \begin{aligned}
\iu A_m^{i}\chi^+  \cdot \sigma_{i} = & \tfrac{1}{2} (\Omega_m)_{\hat a\hat b} \hat \gamma^{\hat a\hat b} \chi^+ \ , \\
\iu F^{mi} \chi^+  \cdot \sigma_i  = & \tfrac{1}{6} \epsilon^{mnpq}F_{\hat a\hat b npq} \hat \gamma^{\hat a\hat b} \chi^+ \ , \\
F^{i} \chi^-  \cdot \sigma_i  = & - \iu \tfrac{1}{48} \e^{\phi}(F_{\hat a \hat b8}+\iu F_{\hat a \hat b9})  \hat \gamma^{\hat a \hat b} \chi^- \ , \\
A_m^0 = & - (\Omega_m)_{89} \ , \\
F^{m0}  = & - \tfrac{1}{6} \epsilon^{mnpq}F_{89 npq} \ , \\
G_{mn}  = & \tfrac18 (H_{mn8}+\e^{\phi}F_{mn9} + \iu H_{mn9} - \iu \e^{\phi}F_{mn8})\ , \\
\tilde G_{mn}  = & \tfrac18 \iu \e^{\phi} (F_{np8}+ \iu F_{np9} )\ ,
\end{aligned} \end{equation}
and we choose the charge conjugation matrix $C_{(6)}$ such that $C_{(6)}= C \hat \gamma_8$ with $C \in {\rm su}(2)$ is the charge conjugation matrix.

In order to reduce the supersymmetry condition \eqref{eq:susy4dN=2} to an $N=1$ form we have to introduce another projector. The simplest ways to do so is by adding another stack of D7-branes. In the case of D3-branes, there are two compatible D7-brane projections that do not introduce any new defects into the four-dimensional worldvolume theory.
We choose the flat indices such that these additional kappa symmetry projectors read
\begin{equation}
\hat \Gamma_{{\rm D}7} = \gamma_{(4)} \hat \gamma_{4567} {\cal P}_1 \ , \qquad \hat \Gamma_{{\rm D}7'} = \gamma_{(4)} \hat \gamma_{4589} {\cal P}_1 \ .
\end{equation}
Acting with these projections on $\hat \varepsilon$ implies that $\eta^{\pm}$ obey
\begin{equation}
\hat \gamma_{4567} \chi^{\pm} = \chi^{\pm} \ , \qquad \hat \gamma_{4589} \chi{\pm} = \chi^{\pm} \ , \qquad \hat \gamma_{6789} \chi^{\pm} = - \chi^{\pm} \ ,
\end{equation}
or, taking into account the chirality of $\chi^+$ and $\chi^-$, this means
\begin{equation}
\hat \gamma_{45} \chi^{\pm} = \mp \iu \chi^{\pm} \ , \qquad \hat \gamma_{67} \chi^{\pm} = \pm \iu \chi^{\pm} \ , \qquad \hat \gamma_{89} \chi^{\pm} = \pm \iu \chi^{\pm} \ .
\end{equation}
Moreover, we can choose charge conjugation such that
\begin{equation}
\Gamma^{468} \chi^- = \chi^+ \ .
\end{equation}
Then we find from commutating \eqref{eq:susy4d} with $\hat \gamma_{45}$, $\hat \gamma_{67}$ and $\hat \gamma_{89}$ that
\begin{equation} \label{eq:susy4dN=1} \begin{aligned}
(\tilde \nabla_m + \iu A_m  + \iu V_n \gamma^n \gamma_m) \eta^+ + M \gamma_m \eta^- & = 0 \ , \\
(\tilde \nabla_m - \iu A_m  - \iu V_n \gamma^n \gamma_m) \eta^- + \bar M \gamma_m \eta^+ & = 0 \ ,
\end{aligned} \end{equation}
where
\begin{equation}\label{eq:susy4dN=1def}
A_m =  (\Omega_m)_{67} + (\Omega_m)_{89} - (\Omega_m)_{45} \ , \qquad V_m = \tfrac{1}{8} \e^{\phi} ( F_m  + F_{m4567} + F_{m4589} - F_{m6789} ) \ ,
\end{equation}
and
\begin{equation}
M  = \tfrac{1}{8} \e^{\phi}( F_{468} +F_{578} + F_{569} - F_{479} + \iu (F_{478} + F_{469} + F_{579} - F_{568} ))  \ .
\end{equation}

If we want to derive the Euclidean versions of the supersymmetry conditions \eqref{eq:susy4dN=4}, \eqref{eq:susy4dN=2} and \eqref{eq:susy4dN=1}, we have to consider an E3-instanton in ten-dimensional spacetime. This changes the kappa symmetry projectors slightly, in that we find the chirality projectors
\begin{equation}
 \gamma_{(4)} \eta^{\pm} = \pm \eta^{\pm} \ , \qquad \hat \gamma_{(6)} \chi^{\pm} = \pm \chi^{\pm} \ .
\end{equation}
Thus we get the Euclidean equations by replacing
\begin{equation}
F_\bullet \to - \iu F_\bullet \ , \qquad \gamma_{(4)} \to -\iu \gamma_{(4)} \ .
\end{equation}
so that \eqref{eq:susy4d} becomes
\begin{equation} \label{eq:susy4dEuclideanN=4} \begin{aligned}
(\tilde \nabla_m + \tfrac12 (\Omega_m)_{ab} \hat \gamma^{ab}  + \tfrac{1}{16} \e^{\phi} ( 2 \iu F_n \gamma^n \gamma_m  \gamma_{(4)}  + \tfrac{1}{6} \iu F_{abnpq} \hat \gamma^{ab} \epsilon^{npqr}\gamma_{r} \gamma_m ))\chi_\alpha^{+} \otimes \eta_\alpha^+ & \\
+(\tfrac18 H_{mna} \hat \gamma^a \gamma^{n} + \tfrac{1}{16} \e^{\phi} (- \tfrac{1}{3} F_{abc} \hat \gamma^{abc} \gamma_{m}
+ 2 F_{amn} \hat \gamma^{a} \gamma^{n}  + \iu  \epsilon_m{}^{npq} F_{anp} \hat \gamma^{a} \gamma_{q} ))\chi_\alpha^{-} \otimes \eta_\alpha^- &  = 0 \ .
\end{aligned} \end{equation}

In the $N=1$ projection let us assume that both stacks of D7 branes fill out the time direction. Then we find
\begin{equation}
\hat \gamma_{0567} \eta^{\pm} = -\iu \eta^{\pm} \ , \qquad \hat \gamma_{0589} \eta^{\pm} = -\iu \eta^{\pm} \ , \qquad \hat \gamma_{6789} \eta^{\pm} = - \eta^{\pm} \ ,
\end{equation}
or
\begin{equation}
\hat \gamma_{05} \eta^{\pm} = \mp \eta^{\pm} \ , \qquad \hat \gamma_{67} \eta^{\pm} = \pm \iu \eta^{\pm} \ , \qquad \hat \gamma_{89} \eta^{\pm} = \pm \iu \eta^{\pm} \ .
\end{equation}
Under these projections, \eqref{eq:susy4dEuclideanN=4} simplifies to
\begin{equation} \label{eq:susy4dEuclideanN=1} \begin{aligned}
(\tilde \nabla_m + \iu A_m  + \iu V_n \gamma^n \gamma_m)\chi^{+} \otimes \eta^+ & \\
 - \tfrac{1}{48} \e^{\phi} F_{abc} \hat \gamma^{abc} \gamma_{m} \chi^{-} \otimes \eta^- &  = 0 \ , \\
 (\tilde \nabla_m - \iu A_m  - \iu V_n \gamma^n \gamma_m)\chi^{-} \otimes \eta^- & \\
 - \tfrac{1}{48} \e^{\phi} F_{abc} \hat \gamma^{abc} \gamma_{m} \chi^{+} \otimes \eta^+ &  = 0 \ ,
\end{aligned} \end{equation}
where
\begin{equation} \label{eq:AVEuclidean}
A_m =  (\Omega_m)_{67} + (\Omega_m)_{89} +\iu (\Omega_m)_{05} \ , \qquad V_m = \tfrac{1}{8} \e^{\phi} ( F_m  -\iu F_{m0567} -\iu F_{m0589} - F_{m6789} ) \ ,
\end{equation}
are now complex quantities.

Note that $\chi^+$ and $\chi^-$ are not related by charge conjugation in six-dimensional spacetime. However, \eqref{eq:susy4dEuclideanN=4} can only be solved if
\begin{equation}\label{eq:MEuclidean}
- \tfrac{1}{48} \e^{\phi} F_{abc} \hat \gamma^{abc} \chi^- = M \chi^+ \ ,
\end{equation}
for some complex $M$. Similarly, we find for
\begin{equation}\label{eq:MEuclidean2}
- \tfrac{1}{48} \e^{\phi} F_{abc} \hat \gamma^{abc} \chi^+ = \bar M \chi^- \ ,
\end{equation}
where (abusing notation) $\bar M$ is not necessarily the complex conjugate of $M$.
In total we find again \eqref{eq:susy4dN=1}, but now the parameters of this equation are determined by (\ref{eq:AVEuclidean}-\ref{eq:MEuclidean2}).

The condition \eqref{eq:susy4dN=1} implies the charged twistor spinor equation that has been studied in \cite{Klare:2012gn} and references therein. The couplings appearing in \eqref{eq:susy4dN=1} combine the auxiliary fields of old and new minimal off-shell supergravity: If we set $M$ to zero, we end up in new minimal supergravity \cite{Sohnius:1981tp,Sohnius:1982fw}. If we set $A$ to zero, we end up in old minimal supergravity \cite{Stelle:1978ye,Ferrara:1978em}. As we know, both formalisms allow for different manifolds to be supersymmetric \cite{Festuccia:2011ws,Dumitrescu:2012ha,Dumitrescu:2012at}.
Thus \eqref{eq:susy4dN=1} must allow for more general solutions than the conditions coming from minimal $N=1$ off-shell supergravity formalism that has been used so far. Indeed, Equation \eqref{eq:susy4dN=1} naturally appears as gravitino variation in 16/16 supergravity \cite{Girardi:1984vq,Lang:1985xk,Siegel:1986sv}, which has exactly the right auxiliary fields\footnote{Note that there is additionally the dilaton as an auxiliary field, which only appears in the equation coming from the dilatino.} to fit \eqref{eq:susy4dN=1} and corresponds to coupling the field theory to the S-multiplet \cite{Komargodski:2010rb}. It would be very interesting to find a classification for the solutions to \eqref{eq:susy4dN=1}. Since the S-multiplet is the most general coupling of field theory to supergravity \cite{Dumitrescu:2011iu}, this should give a full classification of $N=1$ supersymmetric field theories on curved backgrounds.

\section{Supersymmetric curved membranes}
\label{sec:membranes}

Supersymmetric membrane theories are, apart from the classic D3-brane case, particularly interesting as they feature in a prominent role in the construction of many holographic AdS/CFT pairs. In general the worldvolume theory is much less understood and in many cases no perturbative description is even available. However, one can nevertheless study the supersymmetry condition and use methods that do not require a perturbative description.

Note that we will perform the analysis only for the Lorentzian case, but the Euclidean case can of course be easily obtained by Wick rotation.

\subsection{M2-branes}
The kappa symmetry for an M2-brane is \eqref{eq:kappaM2} so that
\begin{equation}\label{eq:M2proj}
\gamma_{(3)} \hat \varepsilon = \hat \varepsilon \ .
\end{equation}
In eleven dimensions there is
\begin{equation}
\Gamma_{(11)}= \tfrac{1}{11!} \epsilon^{M_1 \dots M_{11}} \Gamma_{M_1 \dots M_{11}}= 1 \ ,
\end{equation}
so that this translates into the condition
\begin{equation} \label{eq:M2chiral}
\hat \gamma_{(8)} \hat \varepsilon = \tfrac{1}{8!} \epsilon^{a_1 \dots a_{8}} \hat \gamma_{a_1 \dots a_{8}} \hat \varepsilon =  \hat \varepsilon \ .
\end{equation}
which means that the $G_{\rm R}={\rm SO}(8)$ representation of $\hat \varepsilon$ will be the chiral representation.
We can compute \eqref{eq:anticommkappa} for this projection to be
\begin{equation} \label{eq:susyM2}
(\tilde \nabla_m + \tfrac12 (\Omega_m)_{ab} \hat \gamma^{ab} - \tfrac{1}{12}( \tfrac{1}{4!} G_{abcd} \hat \gamma^{abcd} \gamma_m + \tfrac{1}{4} \epsilon_{m}{}^{np} G_{npab} \hat \gamma^{ab} - G_{mnab} \hat \gamma^{ab} \gamma^{n}) \hat \varepsilon  = 0\ .
\end{equation}

The chiral $G_{\rm R}={\rm SO}(8)$ spinor representation of $\hat \varepsilon$, confer \eqref{eq:M2chiral}, is by triality isomorphic to the fundamental representation of ${\rm SO}(8)$. Thus the decomposition \eqref{eq:spinor_decomposition} actually reads
\begin{equation}
\hat \varepsilon = \sum_{\alpha=1}^8 \chi^\alpha \eta^\alpha \ .
\end{equation}
If we apply this decomposition to \eqref{eq:susyM2}, we find
\begin{equation}\label{eq:susy3d}
(\tilde \nabla_m + \tilde A_m + S \gamma_{m} + \epsilon_{mnp} T^{p} \gamma^{n}) \eta = 0\ ,
\end{equation}
where we defined the ${\rm SO}(8)$ connection $\tilde A_m$ by
\begin{equation}
(\tilde A_m)^\alpha{}_\beta \chi^\beta  = - \tfrac12 ((\Omega_m)_{cd}  - \tfrac{1}{24} \epsilon_{m}{}^{np} G_{npab} )\hat \gamma^{ab}\chi^\alpha \ .
\end{equation}
The other couplings are a traceless symmetric matrix $S$ given by
\begin{equation}
S^\alpha{}_\beta \chi^\beta =  - \tfrac{1}{288} G_{abcd} \hat \gamma^{abcd} \chi^\alpha \ ,
\end{equation}
and three anti-symmetric tensors $T^{m} $ given by
\begin{equation}
(T^{m})^\alpha{}_\beta \chi^\beta = - \tfrac{1}{24} \epsilon^{mnp} G_{npab} \hat \gamma^{ab} \chi^\alpha \ .
\end{equation}

On any manifold there is a simple solution to \eqref{eq:susy3d}, as the maximal holonomy on $\cal S$ is $SU(2)$, which can be easily compensated for by a topological twist, and this twist can preserve up to eight supercharges. However, it is not clear which other supersymmetric theories can be defined on a certain manifold $\cal S$.

So far the formalism is $N=8$. We can project the theory to less supercharges by using 6-branes of M-theory \cite{Hull:1997kt}. The projector in \eqref{eq:Mbrane_exotic} reduces with \eqref{eq:M2proj} to
\begin{equation} \label{eq:M2N=4}
\hat \gamma_{(4)} \chi = \chi \ .
\end{equation}
This projection breaks the Lorentz group of the normal bundle to ${\rm SO}(4)\times {\rm SO}(4)$. Since \eqref{eq:M2N=4} projects to chiral four-dimensional spinors, we find that the R-Symmetry is broken from ${\rm SO}(8)$ to ${\rm SU}(2) \times {\rm SU}(2)$, which is the R-symmetry for an $N=4$ theory. The couplings $S$ and $T$ are accordingly projected to traceless symmetric and anti-symmetric $(4\times 4)$-matrices, respectively. By intersecting two stacks of 6-branes in the M2-brane times two more dimensions, we can further break the R-symmetry to ${\rm U}(1)$, which gives us the $N=2$ formalism in three dimensions.

\subsection{M5-branes}

We will study now a calibrated $M5$ brane inside a supersymmetric M-theory background. This should determine on which manifolds we can put a $(2,0)$ theory in a supersymmetric fashion. Subsequently we will discuss the projection to $(1,0)$ theory. Note that by dimensional reduction to five dimensions, the derived supersymmetry conditions also describe five-dimensional field theories.

In the following we will ignore worldvolume fluxes for the M5-brane, i.e.\ ${\cal H}=0$.
Then the kappa symmetry projection \eqref{eq:kappa_symm} with the kappa symmetry projector \eqref{eq:kappaM5} means that the spinor is chiral in six dimensions
\begin{equation}\label{eq:M5chiral}
\gamma_{(6)} \hat \varepsilon= \hat \varepsilon\ ,
\end{equation}
where we defined $\gamma_{(6)}= \epsilon^{m_{1} \dots m_6 } \tfrac{1}{6!} \gamma_{m_1 \dots m_6}$.
Then from \eqref{eq:anticommkappa} we find
\begin{equation} \label{eq:susyM5}
(\tilde \nabla_m + \tfrac12 (\Omega_m)_{ab} \hat \gamma^{ab} - \tfrac{1}{12}( \tfrac{1}{3!} G_{anpq} \hat \gamma^a \gamma^{npq}{}_m + \tfrac{1}{3!} G_{abcn}  \hat \gamma^{abc} \gamma^{n}{}_m - G_{mnpa} \hat \gamma^{a} \gamma^{np} - \tfrac{1}{3} G_{mabc}  \hat \gamma^{abc} ) )\hat \varepsilon = 0\ ,
\end{equation}
where $\tilde \nabla$ is the induced connection on the worldvolume.

We can decompose the $\hat \varepsilon$ into
\begin{equation}
\hat \varepsilon = \sum_{i=1}^2 \chi^i \otimes \eta^i + \chi^{c\, i} \otimes \eta^{c\,i} \ ,
\end{equation}
where the $(\chi^\alpha)=(\chi^i, \chi^{c\,i})$ together transform in the fundamental of ${\rm USp}(4)$. Then the $(\eta^\alpha)=(\eta^i, \eta^{c\,i})$ are of positive chirality, due to \eqref{eq:M5chiral}.
We can rewrite \eqref{eq:susyM5} as
\begin{equation}\label{eq:susyM5f}
(\tilde \nabla_m + \tilde A + \tfrac{1}{2} T_{mnp} \gamma^{np} + S_n \gamma^n \gamma_m) \eta = 0 \ ,
\end{equation}
where the ${\rm USp}(4)$ connection $\tilde A$ for the topological twist is defined by
\begin{equation}
\tilde A^\beta{}_\alpha \chi^\beta = - \tfrac12 ((\Omega_m)_{ab} + \tfrac{1}{36} \epsilon_{ab}{}^{cde} G_{mcde})\hat \gamma^{ab} \chi^\alpha \ .
\end{equation}
The matrices $T_{mnp}$ are Hermitian and defined by the action
\begin{equation} \label{eq:susyM5def1}
(T_{mnp})^\beta{}_\alpha \chi^\beta = - \tfrac{1}{12} (\tfrac{1}{3!} \epsilon_{mnp}{}^{qrs} G_{aqrs} + 2 G_{amnp}) \hat \gamma^a \chi^\alpha \ .
\end{equation}
The matrices $S_n$ are anti-Hermitian and are elements of ${\rm usp}(4)$, defined by
\begin{equation}\label{eq:susyM5def2}
(S_{m})^\beta{}_\alpha \chi^\beta = \tfrac{1}{144} \epsilon^{abcde} G_{mcde} \hat \gamma_{ab} \chi^\alpha \ .
\end{equation}
This gives the conditions to preserve some supersymmetry within the framework of $(2,0)$ theory.

Also for M5-branes there should be a projection to the $(1,0)$ theory. For that we will consider a stack of M9-branes \cite{Hull:1997kt,Bergshoeff:1998bs} that fills out the M5-brane plus four other directions. Using \eqref{eq:Mbrane_exotic} we therefore find an additional projector $\tilde \gamma_{(10)} = \gamma_{(6)} \hat \gamma_{(4)}$ and, since \eqref{eq:M5chiral} holds, the condition reduces to
\begin{equation}\label{eq:M9chiral}
\hat \gamma_{(4)} \chi= \chi\ .
\end{equation}
This means that the spinors $\chi^\alpha $ are projected to the four-dimensional spinors of positive chirality $(\chi, \chi^c)$ and thus the R-Symmetry group ${\rm USp}(4)$ is broken to ${\rm USp}(2) = {\rm SU}(2)$. This projections affects all couplings in the same way and simply projects $\tilde A$ and $S_m$ onto ${\rm su}(2)$ representations, while $T_{mnp}$ is projected to a scalar
\begin{equation}
\tilde T_{mnp} = - \tfrac{1}{12} (\tfrac{1}{3!} \epsilon_{mnp}{}^{qrs} G_{10qrs} + 2 G_{10mnp}) \ .
\end{equation}
Equation \eqref{eq:susyM5f} does not change otherwise.

\section{Examples}
\label{sec:examples}
We want to give here some non-trivial examples of calibrated branes in flux backgrounds. The simplest known flux backgrounds are compactifications to anti-de Sitter (AdS) spacetimes.
If we consider ${\rm AdS}_5 \times S^5$, the supersymmetry of the ten-dimensional background is supported by five-form flux on $S^5$. We see from \eqref{eq:susy4dN=4def} and \eqref{eq:susy4dN=1def} that the (self-dual) $F_5$ flux enters the D3-brane worldvolume supersymmetry condition if it fills out an odd number of directions.
Indeed, this additional term in \eqref{eq:susy4dN=4} and \eqref{eq:susy4dN=1} can cancel the covariant derivative of a spinor on a sphere or on AdS spacetime \cite{Lu:1998nu}.
Thus a stack of D3-branes wrapping an $S^3 \subset S^5$ (and being a point particle at the center of AdS$_3$) or wrapping an $S^1 \subset S^5$ (and filling out ${\rm AdS}_3$) can be made supersymmetric. The corresponding four-dimensional field theory has a non-trivial $U_m$ (or $V_m$ in the $N=1$ case) and therefore can live on $S^3$ and ${\rm AdS}_3$, respectively.
These examples also exist if we replace $S^5$ by some five-dimenional Sasaki-Einstein manifold. It might be that in that case the theory is in addition topologically twisted.

Similarly, we can place M2- and M5-branes in the holographic backgrounds ${\rm AdS}_7 \times S^4$ and ${\rm AdS}_4 \times S^7$ and find similar three- and six-dimensional supersymmetric field theories. When the M2-brane  completely fills out ${\rm AdS}_3 \subset {\rm AdS}_7$ or $S^3 \subset S^7$, it automatically gives a supersymmetric field theory. Similarly, if an M5-brane fills out ${\rm AdS}_5 \times S^1$ or sits at the center of ${\rm AdS}_4$ and wraps an $S^5\subset S^7$, we find supersymmetric theories. Also here we could discuss the general Sasaki-Einstein case.

Also in type IIB compactifications on conformal Calabi-Yau spaces in the spirit of \cite{Giddings:2001yu} we have calibrated four-cycles in the internal Calabi-Yau, and these four-cycles are usually conformally K\"ahler. In the same spirit as in the discussion of AdS backgrounds, here the five-form flux leads to additional couplings on the world volume theory that cancel the warp factor in \eqref{eq:susy4dN=1}. An additional topological twist with U(1) makes the four-dimensional field theory is supersymmetric on any conformally K\"ahler manifold. A related discussion of E3-instantons in F-theory setups has been studied in \cite{Bianchi:2011qh,Bianchi:2012kt,Martucci:2014ema}. The discussion gets even more interesting in the case where supersymmetry is broken by fluxes on the classical level and only restored by non-perturbative effects, suggesting similar non-perturbative effects to take place on the brane world-volume, cf.\ \cite{Bianchi:2012kt}.

\section{Discussion}
\label{sec:conclusion}

In this work we showed how calibrated branes in a supersymmetric flux background give rise to supersymmetric field theories on curved manifolds. In particular we derived a necessary condition that determines whether a certain curved manifold can even admit supersymmetric branes, and therefore supersymmetric field theories. We did not only recover the well-known topological twist, but also linked the various other terms to fluxes in the ambient geometry. The supersymmetry condition of a single stack of branes naturally has an $N=4$ form, and imposing suitable projections we could also derive the $N=2$ and $N=1$ conditions.

It has been known for a long time that $N=2$ and $N=4$ theories can be defined on any manifold, by using the topological twist \cite{Witten:1994ev}. However, depending on the manifold there seem to be various other ways to define supersymmetric theories, without the need for the theory to be topological.
It would be very interesting to study the structure of supersymmetric field theories that can live on a given manifold.

In the $N=1$ case we made contact with the similar supersymmetry conditions coming from coupling the theory to off-shell supergravity and showed that both old and new minimal supergravity can only describe subcases, while the most general case should be described by coupling to off-shell 16/16 supergravity. More precisely, the $N=1$ supersymmetry condition in four dimensions derived in this work allows for all the couplings coming from auxiliary fields of the S-multiplet discussed in \cite{Komargodski:2010rb}. Since this is the most general way to couple a field theory to supergravity \cite{Dumitrescu:2011iu}, we suspect that studying the solutions to this supersymmetry condition might lead to a classification of supersymmetric field theories on curved backgrounds.
This condition still fulfills the charged conformal spinor equation as observed in \cite{Klare:2012gn}. It would be very interesting to find the general solution to this equation, as it might lead to a classification for manifolds that support $N=1$ supersymmetric field theories.

We performed a similar analysis for membranes in M-theory and derived the conditions to make three-dimensional field theories and six-dimensional $(2,0)$ and $(1,0)$ theories supersymmetric on a curved background. Dimensional reductions of the six-dimensional theories should also lead to the condition for five-dimensional field theories on a curved background to be supersymmetric.

Many holographic pairs originate from a stack of D3-branes or M-theory membranes. The results presented here indicate that many supersymmetric field theories on curved manifolds can originate from such stacks of branes, and therefore should have a holographic dual. The identification of couplings in the worldvolume supersymmetry condition with bulk form field strengths also indicates the flux configurations that a gravitational background dual to a given supersymmetric field theory on a curved manifold should have. By studying \eqref{eq:embedding} in the neighborhood of the brane one can then construct a ten- or eleven-dimensional supersymmetric background that the brane embeds into, and thereby also a holographic dual to the field theory on the brane.

As pointed out in this work the analysis can be performed for calibrated cycles of arbitrary dimension. It would actually be interesting to understand the geometry of calibrated cycles in arbitrary flux backgrounds. The tools developed in this work might help in initiating such an analysis.

We did not study punctures of any kind in this paper. They can easily be included in the setup by non-trivially intersecting the brane worldvolume with other brane stacks. These punctures usually further break supersymmetry at their position, and also manifestly break Lorentz invariance of the worldvolume.
We also did not discuss worldvolume fluxes, as they considerably complicate the analysis by modifying the kappa symmetry condition. Moreover they break worldvolume Lorentz-invariance.
It would be very interesting to study supersymmetry conditions in the presence of worldvolume fluxes and punctures.

\section*{Acknowledgments}
We thank I.~Bah, F.~Benini, C.~Closset, S.~Ferrara, R.~Minasian, A.~Tomasiello, C.~Vafa and A.~Zaffaroni for useful discussions. We would give our special thanks to the Department of Physics and Astronomy at University of Southern California in general and to I.~Bah and his family in particular for hospitality during the initial phase of this project. We are very grateful to C.~Klare and N.~Mekareeya for comments on the final draft.

\end{document}